# An extended scenario for the Schrödinger equation

M. Çapak, Y. Cançelik, Ö. L. Ünsal, Ş. Atay and B. Gönül

Department of Engineering Physics, University of Gaziantep, 27310, Gaziantep-Turkey

**Abstract**

The concept of the elegant work introduced by Lévai in Ref. [5] is extended for the solutions of the Schrödinger equation with more realistic other potentials used in different disciplines of physics. The connection between the present model and the other alternative algebraic technique in the literature is discussed.

## 1. INTRODUCTION

A simple method of investigating the solution of the Schrödinger equation, which is related to the work of Bhattacharjie and Sudarsan [1] has been known for a long time. These authors applied their method to the hypergeometric, confluent hypergeometric and Bessel equations. Later it turned out that it can be related to algebraic techniques of solving differential equations [2]. Another systematic application of this method (to the hypergeometric functions) has been carried out by Natanzon [3] independently. In the following years, there has been also renewed interest in simple quantum mechanical systems as a result of the introduction of two important concepts: supersymmetric quantum mechanics (SUSYQM) and shape invariance. For a comprehensive review on this topic, the reader is referred to [4] and the related references therein. In the light of this progress and the previous works mentioned, a significant question has then arisen regarding if there are any other special functions which are solutions of the Schrödinger equation with shape invariant potentials. This question has been answered in detail by Lévai [5] through the consideration of the link between the works in [1-3] and the formalism of SUSYQM, deducing a condition which has to be satisfied by any special function leading to the orthogonal polynomials and exactly solvable shape invariant potentials. Besides the results obtained, the combination of SUSYQM with traditional approaches to solvable potentials proved to be fruitful. For instance, Refs. [6-15] involves some applications of the original idea discussed in [5].

However, to our knowledge, this formalism up to now has been used only to study exactly solvable systems. Therefore, it needs a meticulous modification to also solve more realistic other systems as the ones of interest in this article. Within this context the main motivation behind the present work, bearing in mind that realistic physical problems can practically never be solved exactly, is to suggest a more comprehensive and generalized model using the spirit of the investigation in [5], which escaped notice in other publications. As an illustration, the present novel scheme is applied first to quartic anharmonic oscillator since there has been a great deal of interest in anharmonic oscillators due to their phenomenological as well as methodological use in physics. These potentials also has the characteristics of being a rather simple model where many non-trivial features essential to understanding quite complicated system may be implemented. Their exact solutions however for arbitrary couplings are hard to find. This has culminated into the development of many fascinating techniques based on perturbative and non-perturbative approaches, for a recent review see [16]. Thus, it appears challenging to test our formalism in avoiding the failure of other perturbation series for the treatment of the quartic anharmonic oscillator. For completeness, the model proposed will also be applied to the well-known sextic oscillator problem, which provides an alternative perspective in justifying the capability of widespread applicability of the present scheme.

Furthermore, as theoretical description of energy-dependent interactions have been subjected to intensive investigations during the last decades and the use of such phenomenological potentials in wave equations proved useful in dealing with problems in atomic, molecular as well as nuclear and particle physics [17-21], the second piece of the application section is devoted to such interactions. In particular, the presence of energy-dependent contribution in the potential has several implications modifying the usual rules of quantum mechanics. To get an insight into the clean route visualizing such modifications within the frame of the new scheme, linear energy dependency are considered which has not been previously studied under the traditional models discussed above. This work provides a benchmark test for the present model calculations, too, as the full result obtained within the new formalism should reduce to the familiar solutions concerning with the exactly solvable energy-independent potentials in case the potential parameter related to the energy vanishes.

In section 2 we present the formalism of our method, and in section 3 we apply it to distinct cases including different potentials in order to convince the reader regarding the reliability and flexibility of the model introduced. Section 3 also discusses a significant result behind the calculations and clarifies the inter-relation between the present formalism and the one used in [16] that was performed within the frame of an extended SUSYQM theory. Finally, concluding remarks are given in section 5.

## 2. FORMALISM

It is well known that the general framework of non-relativistic quantum mechanics is by now well understood and its predictions have been carefully proved against observations. Physics is permanently developing in a tight interplay with mathematics. It is of importance to know therefore whether some familiar problems are of particular case of a more general scheme or to search if a map between the radial equations of two different scenarios exist. It is hence worthwhile to devote ourselves to the clarification of this point through the rest of this article.

Considering the Schrödinger equation ($\hbar = 2m = 1$)

$$\frac{d^2\Psi}{dx^2} + (E - V(x))\Psi = 0 \qquad (1)$$

we suggest, for a generalized formalism, that

$$\Psi(x) = f(x)F(g(x))h(x) \quad , \qquad (2)$$

where $f(x)F(g)$ yields an algebraic closed solution for exactly solvable potentials [5-15] with $F(g)$ being a special function which satisfies a second-order differential equation

$$\frac{d^2F}{dg^2} + Q(g)\frac{dF}{dg} + R(g)F(g) = 0 \quad , \qquad (3)$$

while $h(x)$ is the moderating function in connection with a perturbing piece of the full potential corresponding to (2). The form of $Q(g)$ and $R(g)$ is already well defined for any special function $F(g)$ when dealing with analytically solvable potentials. However, in case of the consideration of a realistic non-exactly solvable problem one should derive reliable expressions, in an explicit form, for plausible definitions of the related $Q(g)$ and $R(g)$. This is the significant point in the framework of the new formalism to reach physically meaningful solutions.

Substituting Eq. (2) into (1) leads to

$$\frac{d^2F}{dg^2} + \frac{dF}{dg}\left(\frac{g''}{(g')^2} + \frac{2f'}{fg'} + \frac{2h'}{hg'}\right) + F(g)\left(\frac{f''}{f(g')^2} + \frac{2f'h'}{fh(g')^2} + \frac{h''}{h(g')^2} + \frac{E - V(x)}{(g')^2}\right) = 0 \quad . \qquad (4)$$

From the comparison of Eqs. (3) and (4) it follows that

$$Q(g(x)) = \frac{g''}{(g')^2} + \frac{2f'}{fg'} + \frac{2h'}{hg'} \quad (5)$$

and

$$R(g(x)) = \frac{f''}{f(g')^2} + \frac{2f'h'}{fh(g')^2} + \frac{h''}{h(g')^2} + \frac{E - V(x)}{(g')^2} \quad . \quad (6)$$

Obviously, Eqs. (4-6) reduce to Eqs. (3.4-3.6) in Ref. [5] for the consideration of exact solvability, in which case $h(x)$ in the equations above goes to a constant value. Gaining confidence from this observation we proceed with

$$V(x) = V_{ES}(x) + \Delta V(x)$$
$$E = E_{ES} + \Delta E \quad (7)$$

in accordance with our choice in (2), which means that potentials considered in this article are admitted as the sum of an exactly solvable potential with a perturbation or a moderating piece. Hence, the aim in this perspective is to reveal the corrections to energy ($\Delta E$) and wavefunction $h(x)$ for a given $\Delta V(x)$, as the main piece of the solutions leading to exact solvability can easily be found from the literature.

The use of (7) within Eq. (6) produces coupled equations in the form of

$$E_{ES} - V_{ES}(x) = R_{ES}(g(x))(g')^2 - f''/f \quad , \quad (8)$$

and

$$\Delta E - \Delta V(x) = \Delta R(g(x))(g')^2 - 2f'h'/(fh) - h''/h \quad (9)$$

where $R_{ES}(g) + \Delta R(g)$ should certainly reproduce Eq. (6). Similarly, Eq. (5) can be decomposed as

$$Q_{ES}(g(x)) = g''/(g')^2 + 2f'/fg' \quad , \quad \Delta Q(g(x)) = 2h'/hg' \quad \Rightarrow \quad Q = Q_{ES} + \Delta Q \quad . \quad (10)$$

To be more practical it is reminded that $f''/f = (f'/f)^2 + (f'/f)'$ and the same is valid for $h''/h$ in the equations above, which transform Eqs. (8) and (9) into more applicable forms

$$E_{ES} - V_{ES}(x) = \frac{g'''}{2g'} - \frac{3}{4}\left(\frac{g''}{g'}\right)^2 + (g')^2\left[R_{ES}(g(x)) - \frac{1}{2}\frac{dQ_{ES}}{dg} - \frac{1}{4}Q_{ES}^2(g(x))\right] \quad , \quad (11)$$

and

$$\Delta E - \Delta V(x) = -\left(\frac{g''}{2} + \frac{f'g'}{f}\right)\Delta Q(g(x)) + (g')^2\left[\Delta R(g(x)) - \frac{1}{2}\frac{d(\Delta Q)}{dg} - \frac{1}{4}\Delta Q^2(g(x))\right] . \quad (12)$$

The result of this brief investigation opens a gate to the reader for the visualization of the explicit form of the correction ($\Delta E$) to the energy. Unfortunately, there seems a problem naturally arised in calculating the correction term owing to the presence of two unknown: $\Delta Q(g)$ and $\Delta R(g)$ on the right hand side of Eq. (12). To circumvent the resulting drawback and proceed safely we need to go back Eq. (4) and substitute the definitions given by (7) in it, which leads us to handle

$$\Delta E - \Delta V(x) = -\frac{2h'}{h}\left(\frac{f'}{f} + \frac{F'(g(x))g'}{F(g(x))}\right) - \frac{h''}{h} \qquad (13)$$

that is another form of (9). Thus, equating (9) and (13) and remembering the form of $\Delta Q$ in Eq. (10) we arrive at

$$\Delta R(g) = -\Delta Q(g)\frac{F'(g)}{F(g)} \qquad (14)$$

which is vital to overcome the problem encountered in (12). As $F(g)$ is well defined for a given exactly solvable potential, evidently one needs here to find only an appropriate expression for $\Delta Q(g)$ to be employed in (12) that reveals clearly the full solution. However, singular functions appearing in Eqs. (13-14), and subsequently in (12), are systematically generated when dealing with excited state wavefunctions of any given potential due to the zeros of $F(g)$ function. The effects of this consideration on the calculations are discussed in section 3.2.

Before closing this section, we should remark that once choosing carefully $Q_{ES}(g)$ and $R_{ES}(g)$ for the analytically solvable part ($V_{ES}(x)$) of the full potential under investigation we can easily set a proper internal function $g(x)$ and considering Eq. (5)

$$f(x) \approx (g')^{-1/2}\exp\left[\frac{1}{2}\int^{g(x)}Q_{ES}(g)dg\right], \qquad (15)$$

as discussed in Ref. [5], which are used in (12) to find corrections to the solutions of the exactly solvable piece.

The application of the model to specifically chosen different potentials is discussed in the following section.

## 3. APPLICATION

Special care has to be taken in the application of the model as the results obtained are crucial in the interpretation of the system behaviour in terms of the Hamiltonian described in this work. To reveal especially the flexibility of the scheme used particular cases are discussed below.

### 3.1. Quartic Oscillator

In the light of experiences gained from successful modeling based on anharmonic oscillators, an obvious step in the direction of improvement is to define modifications more accurately brought by anharmonic terms leading to more precise descriptions of the systems considered.

Keeping this point in mind, and also to clarify the relationship between the procedure proposed in this article and the one [16] in the literature, together with the comparison of the results obtained, we restrict ourselves to the Schrödinger equation in one dimension ($\ell = 0$) and consider the anharmonic potential as

$$V(x) = x^2 + \beta x^4 , \tag{16}$$

in which the first piece $V_{ES}(x) = x^2$ represents the well-known exactly solvable harmonic oscillator potential. From the differential equation of the Hermite polynomials [22] one can see that

$$F(g) = \exp(-g^2/2) H_n(g) \quad , \quad R_{ES}(g) = 2n+1-g^2 \quad , \quad Q_{ES}(g) = 0 \quad , \quad g(x) = a^{1/2} x \tag{17}$$

where $a (= w/2)$ is the parameter related to $E_{ES}$. Clearly, from Eqs. (11) and (15), the main contributions through to the closed analytical solutions of the harmonic oscillator are

$$E_{ES} = 2a(n + \frac{1}{2}) \quad , \quad n = 0,1,2,... \quad , \quad \Psi_{ES} = f(x) F(g) = \exp(-\frac{g^2}{2}) H_n(g) \tag{18}$$

in which $\Psi_{ES}$ is the unnormalized wavefunction for the exactly solvable piece of the unharmonic oscillator.

As the whole potential in (16) has no analytical solution, one may expand the related functions in terms of the perturbation such that $\Delta V(r) = \sum_{N=1}^{\infty} \Delta V_N(r)$ and $\Delta \varepsilon_n = \sum_{N=1}^{\infty} \varepsilon_{nN}$ where $N$ denotes the perturbation order. In connection with this idea we choose, after some exhaustive analyses, the form of $\Delta Q$ as

$$\Delta Q(g) = -\frac{2}{g'} \sum_{N=1}^{\infty} j_N x^{2N+1} \tag{19}$$

and substitude all the above expansions into Eq. (12) by equating terms with the same power of the perturbation order on both sides, which yield the modifying terms in the frame of coupled equations at successive orders for different states. It is stressed that as $g(x)$, $f(x)$, $F(g)$ and finally $\Delta R(g)$, from Eqs. (14) and (19), are known one can compute readily the corrections to the whole solution using (12) at each perturbation order for a quantum state of interest. Before discussing the calculation technique of the corrections to the energy, it is reminded that the modifying function in Eq. (2) is formed consistently as

$$h(x) = \exp\left(\frac{1}{2}\int \Delta Q dg\right) \qquad (20)$$

as a consequence of the choice in (19) and the eventual use of it in (10).

The systematic calculation of energy corrections in different orders involving large $N-$ values offers no difficulty if we resort a computer algebra system like Mathematica. The repeat of our calculations for large successive orders reproduces similar relations in a manner of equation hierarchy. This realization leads us to generalize anharmonic oscillator solutions obtained within the frame of (12), without solving the Schrödinger equation. To calculate the energy values individually at each order we need to solve

$$\sum_{k=0}^{N} j_k j_{N-k} - \beta \delta_{N1} = 0 \qquad (21)$$

in which $\delta$ is the Kronecker delta and $j_0 = a = w/2$. The perturbation coefficients above can be computed through

$$j_N = (2N + 2n + \alpha_n)^{-1}\left(\sum_{k=0}^{N-1} j_k j_{N-k-1} - \delta_{N1} - \beta \delta_{N2}\right) \qquad (22)$$

where $\alpha_n = (n-1) + \alpha_{n-1}$ for the excited states ($n \geq 1$) and $\alpha_0 = 1$ in the case of ground ($n = 0$) and first excited state ($n = 1$) calculations. As a matter of fact, the only data that are needed when using Mathematica is (22) to solve (21) producing energy values through the perturbation orders for any quantum state.

The calculations are carried out for different range of $\beta-$ values and the results obtained for different states at various orders are compared to those of the work in [16]. The agreement is remarkable in the whole range of $\beta-$ values. All the numerical results produced by completely different mathematical procedures of the two alternative approaches, the present one and the other in [16], are exactly the same, which for clarity are not repeated here. This interesting coincidental outcome is of course due to the natural inheritance of the same calculation scheme, Eqs. (21) and (22), in both model. As the same results tabulated in [16] through the Tables (1) and (6) are appeared naturally in the present work with the same precision, and also the accuracy, convergency and the success of the identical model are well discussed in [16] when compared to other techniques available in the related literature, we intend in this section to focus our attention only to this interesting inter-connection between the seemingly alternative but in fact identical prescriptions for the treatments of bound states in non-relativistic domain of the subatomic world.

The most significant piece in [16] is Eq. (8) to find energy corrections through the model used,

$$\Delta E - \Delta V(x) = -\left[\Delta W_n^2(x) + 2W(x)\Delta W(x)\right] + \Delta W'(x) \tag{23}$$

where $W(x)$ and $\Delta W(x)$ are the superpotentials, concerning with the exactly solvable part $V_{ES}(x)$ and the perturbing piece $\Delta V(x)$ respectively, as appeared correspondingly in (16) above. From the extended definitions of superpotential terms in Ref. [16] by employing the spirit of the standard treatment of SUSYQM, we make clear that

$$W_n(x) = -\frac{d}{dx}\ln \Psi_n^{ES} = -\left(\frac{f'}{f} + \frac{F_n' g'}{F_n}\right) \quad , \quad \Delta W(x) = -\frac{h'}{h} \tag{24}$$

Certainly, the substitution of (24) into (23) yields Eq. (13) which can easily be transformed to Eqs. (9) and subsequently (12) as discussed in the previous section, clarifying the reason behind obtaining the identical results. Further, from the definitions of $\Delta W$ in (24) and $\Delta Q$ in (10) and also (20) one can find an explicit relationship such that $\Delta Q = -2\Delta W/g'$ which makes another link between the theoretical considerations of the models being analysed in this section.

Afterall, this brief but concrete analysis sheds a light on a remarkable coincidence regarding the identical treatment of the two alternative scenarios underlined. This investigation also completes the idea of Lévai [5] in which he has related his simple analytic scheme with the treatment procedure in the standard SUSYQM, as the present discussion has made clear the close relation between the generalized work introduced in this article and the method proposed in [16] within the extended framework of SUSYQM, in a similar but extended manner used by [5].

## 3.2. Sextic Oscillator

To improve the precision of the description of bistable systems one has to add a sextic term to the quadratic anharmonic oscillator equation discussed above. Though this section deals with the applications involving general form of sextic oscillators, we need first to remind briefly a peculiar behaviour of such potentials in case it is quasi-exactly solvable, which would be useful in understanding the mathematical procedure behind the present calculations leading to the energy values in case the sextic oscillator potential of interest is non-solvable.

The quasi-exactly solvable form of sextic oscillator potentials with a centrifugal barrier is defined [23]

$$V(x) = \frac{(2s-1/2)(2s-3/2)}{x^2} + \left[a^2 - 4b\left(s+\frac{1}{2}+M\right)\right]x^2 + 2abx^4 + b^2x^6 \qquad (25)$$

where $x \in [0,\infty)$ and $M$ is a non-negative integer. For any value of $M$, leading to certain combinations of potential parameters, only $M+1$ solutions for the related Schrödinger equation can be obtained in an algebraic fashion. The simplest solutions are obtained for $M=0$ and $M=1$.

Starting with $M=0$ case and considering Eq. (7),

$$V_{ES}(x) = a^2x^2 + \frac{(2s-1/2)(2s-3/2)}{x^2} \quad , \quad \Delta V(x) = -4b\left(s+\frac{1}{2}\right)x^2 + 2abx^4 + b^2x^6 \quad , \qquad (26)$$

where the exactly solvable piece, in general, requires

$$F(g) = L_n^{2(s-\frac{1}{4})}(g) \quad , \quad Q_{ES}(g) = (2s-g)/g \quad , \quad R_{ES}(g) = n/g \quad , \qquad (27)$$

in which $g = ax^2$ that yields $f \sim x^{2\left(s-\frac{1}{4}\right)} \exp(-ax^2/2)$. Hence the corresponding ground state algebraic solutions for $V_{ES}(r)$ are

$$E_{ES}^{n=0} = 4as \quad , \quad \Psi_{ES}^{n=0}(x) \sim x^{2(s-1/4)} \exp(-ax^2/2) \,. \qquad (28)$$

To obtain the modifying terms to the solutions in (28), due to the additional term ($\Delta V$) in (26), we set $\Delta Q$ as

$$\Delta Q(g) = -\frac{b}{a^2}g = -\frac{b}{a}x^2 \quad , \qquad (29)$$

and the substitution of which into Eq. (12) reveals that

$$E_{n=0}^{M=0} = 4as \quad , \quad \Psi_{n=0}^{M=0}(x) = f(x)\,h(x) \sim x^{2(s-1/4)} \exp(-\frac{b}{4}x^4 - \frac{a}{2}x^2) \,. \qquad (30)$$

Obviously, the solutions reduces to the analytically solvable harmonic oscillator for the choice of *b=0*, which clarifes that the contributions to $E_{ES}^{n=0}$ due to the two pieces of $\Delta V$ in (26), having opposite signs, cancel each other.

However, the situation for the case of *M=1* is different. Because, the generalized Laquerre polynomial now is not constant, which appears as $F(g) = L_{n=1}^{2(s-\frac{1}{4})}(g) = 2s - ax^2$. Moreover, the change in the potential parameter of the harmonic oscillator like term forces us to re-consider the structure of $g = ax^2$ which now should be $\lambda(a,b,s)x^2$ due to the presence of anharmonic terms in the potential. This behaviour thus requires the replacement of $L_{n=1}^{2(s-\frac{1}{4})} = 2s - ax^2$ with an appropriate another orthogonal polynomial $P_{n=1}^{2(s-\frac{1}{4})}(g) = 2s - \lambda x^2$. With this new consideration the full wavefunction for the first excited state becomes

$$\Psi_{n=1}^{M=1}(x) \sim \left(1 - \frac{\lambda(a,b,s)}{2s}x^2\right) x^{2(s-1/4)} \exp(-\frac{b}{4}x^4 - \frac{a}{2}x^2) \quad , \tag{31}$$

which guides us to use the exact treatment, $\frac{\Psi''}{\Psi} = V - E$, unlike the ground state case, that produces the related energy value as

$$E_{n=1}^{M=1} = 4(as + \lambda) \quad , \quad \lambda_{\mp}(a,b,s) = \frac{1}{2}(a \mp \sqrt{a^2 + 8bs}) \quad . \tag{32}$$

As stated in Ref. [23], $\lambda_{\mp}$ choice has to be made for *n=0* and *n=1* state calculations, respectively. Note that *b=0* case causes $\lambda_{+} \to a$, subsequently $P_{n=1}^{2(s-\frac{1}{4})} \to L_{n=1}^{2(s-\frac{1}{4})}$ which reproduces the known solutions of the usual harmonic oscillator problem.

It has to be finally remarked that the solutions for *M=0* and *M=1* belong to different sextic potentials if $s\left(=\frac{\ell}{2}+\frac{3}{4}\right)$ is the same, as the coefficient of the quadratic term is different then. This shifting in the parameters defines the corresponding energy value for different considerations which are certainly related to the same subsequent perturbation order solutions in distinct quantum states if one deals with non-solvable sextic oscillator problems discussed below.

To complete the discussion in this section, we consider now a general form of the sextic potential in one dimension

$$V(x) = \mu x^2 + \sigma x^4 + \eta x^6 \quad, \tag{33}$$

and solve the corresponding Schrödinder equation approximately within the frame of the present scheme. In this case, Eqs. (21) and (22) become

$$\sum_{k=0}^{N} j_k j_{N-k} - \sigma \delta_{N1} - \eta \delta_{N2} = 0$$

$$j_N = (2N + 2n + \alpha_n)^{-1} \left( \sum_{k=0}^{N-1} j_k j_{N-k-1} - \mu \delta_{N1} - \sigma \delta_{N2} - \eta \delta_{N3} \right) \tag{34}$$

for the systematic calculations of the energy corrections concerning with the quadratic and sextic pieces in (33), where $\alpha_n$ discussed in the previous section. For clarity, as the details of the similar calculation produre for the quadratic potential were well discussed in Ref. [16] through Hermite polynomials using, although indirectly, the same $\Delta Q(g)$ and $\Delta R(g)$ expressions appeared in Eqs. (19) and (14), we illustrate only our application results in Tables 1 and 2.

The agreement is remarkable in the whole range of the potential parameters in the low-lying states. Similar accuracy is observed for the higher quantum levels. Nevertheless, when dealing with excited states the present approach becomes rather cumbersome because the zeros of the wavefunction have to be taken into account explicitly. As expected, due to the consequence of the radial nodes in - more specifically - $F(g)$ and subsequently $\Delta R(g)$ in Eqs. (14) and (12), the present formulae gives small accuracy for large quantum numbers since the perturbation becomes more important.

A question now arises about the convergence of the method just described. Since it is closely related to perturbation theory, as discussed in Section 3.1, one expects it to be asymptotic divergent. Our numerical results confirm this assumption. For some of the potential parameters, in particular the ones chosen in Table 1, the concerning upper root of $j_N = 0$ oscillates about the actual eigenvalue as $N$ increases. The amplitude of the oscillation decreases, reaches a minimum value corresponding lower bounds, and then increases to the upper bounds. Beyond a specific large value of $N$, depending on the energy level of interest, random results are obtained though they remain quite close to the true eigenvalue, as discussed in earlier similar works [28, 29]. Although divergent the present method is stil useful because it certainly improves the perturbation series largely. The most accurate results is obtained from the $N$-value corresponding to the smallest oscillation amplitude. However, the root for the largest $N$ before the oscillation takes place is a quite accurate estimate of the eigenvalue. Such an accuracy cannot be obtained from the perturbation series. Moreover, the present calculations converged quickly for the larger potential parameters shown in Table 2 and reproduce reasonable numerical results for the lower quantum states.

### 3.3. Energy-dependent Potentials

Considering an ongoing belief that standard techniques for approximating a given potential with a separable potential are only applicable to energy independent potentials, a significant extension of the model applications is achieved in this section by incorporating such conventional considerations to those accomodating explicit energy dependence in potentials (EDP) with emphasis on power-law potentials as examples admitting analytical solutions. Heavy quark systems in particular constitute a natural domain for the application of such interactions. Comparing the results of EDP with those of conventional potentials the new features appeared [17-21] in a deep understanding of the systems in high energy physics. For instance, it is now clear that the energy dependent component in the potential has a significant influence on the calculated observables of charmonium and bottomium, unlike the conventional ones. Also the existence of analytical solutions presents a good opportunity in tackling such problems [21].

However, the physical discussion behind EDP applications is not our goal at the present stage. The real question is to know here if there is a failure in the application of our extended formalism to the systems involving EDP, which is the subject of the next section. For the sake of simplicity, we assume a spherical symmetry and a linear energy dependence in the two illustrative examples discussed below.

### 3.3.1. harmonic oscillator

Solutions of such equations with EDP exhibit properties quite unusual with respect to the known solutions of the ordinary Schrödinger equation for the same potential shape. This is particularly spectacular in the case of harmonic oscillator with a linear energy dependence, which is well discussed in [17,18].

The related reduced radial wave equation reads

$$\frac{\Psi''_{n\ell}(x)}{\Psi_{n\ell}(x)} = \left[ V(x, E_{n\ell}) + \frac{\ell(\ell+1)}{x^2} \right] - E_{n\ell} \tag{35}$$

where

$$V(x, E_{n\ell}) = V_{ES}(x) + \Delta V(x, E_{n\ell}) = \left( \frac{w^2 x^2}{4} + \frac{\ell(\ell+1)}{x^2} \right) + \frac{w^2 x^2}{4} (\gamma E_{n\ell}) . \tag{36}$$

The familiar solutions of the energy independent piece of the spherical harmonic oscillator potential $V_{ES}(x)$ in three dimension are

$$E_{ES} = (2n + \ell + \frac{3}{2}) w \quad , \quad \Psi_{ES}(x) = f(x) F(g(x)) \sim g^{(\ell+1)/2} \exp(-g/2) L_n^{(\ell+\frac{1}{2})}(g) \tag{37}$$

due to the choice of the generalized Laguerre polynomials $F(g) = L_n^{(\ell+\frac{1}{2})}$ which leads us to consider $Q_{ES} = (\ell - g + 3/2)/g$ and $R_{ES} = n/g$ in dealing with the algebraic solutions of the corresponding differential equations. This ends up with the strict definitions of the interval functions such as $g = \frac{w}{2} x^2$ and $f = (2w)^{-1/4} g^{(\ell+1)/2} \exp(-g/2)$. Obviously, the substitution of these findings in (8) or (11), as previously discussed in section 3.1, reproduces Eq. (37). Though this reveal anything new, it would be helpful in arriving at the modification terms in their explicit form for understanding the influence of energy dependence of the interaction potential.

From the expertise gained in the analyses of the quartic anharmonic oscillator problem, we can safely set

$$\Delta Q = 1 - \sqrt{1 + \gamma E_{n\ell}} \tag{38}$$

to obtain certain expressions for the corrections brought by the energy component of the potential. It is noted that (38) dies away in the case of $\gamma = 0$, from which Eqs. (14) and subsequently (12) vanish. This confirms the reliability of the choice in (38) which is the key point for benchmark tests when compared to the solutions in connection with the conventional energy independent potentials. The use of (38) in Eqs. (14) and then (12) reveals the modifying term as

$$\Delta E = (\sqrt{1 + \gamma E_{n\ell}} - 1)(2n + \ell + \frac{3}{2}) w \tag{39}$$

for the energy, and similarly one can combine the form of $\Delta Q$ in Eq. (10) with (38) to get

$$h(x, E_{n\ell}) = \exp \left[ \frac{w x^2}{4} (1 - \sqrt{1 + \gamma E_{n\ell}}) \right] \tag{40}$$

for the correction to the wavefunction of the entire system that takes, from Eq. (2), its final form as

$$\Psi(x, E_{n\ell}) = \Psi_{ES}(x) h(x, E_{n\ell}) = C_{n\ell} x^{\ell+1} \exp(-\frac{wx^2}{4}) \exp\left[\left(\frac{wx^2}{4}\right)(1-\sqrt{1+\gamma E_{n\ell}})\right] L_n^{(\ell+\frac{1}{2})}(g) \quad (41)$$

where $g = \frac{wx^2}{2}\sqrt{1+\gamma E_{n\ell}}$ in this case.

Proceeding with (39) to observe the structure of the full energy spectra, we have

$$E_{n\ell} = E_{ES} + \Delta E = (2n+\ell+\frac{3}{2})w\sqrt{1+\gamma E_{n\ell}} \quad . \tag{42}$$

The nonlinear character of the wave equation in (35) is seen explicitly in the above equation. It thus results in a quadratic equation for the eigenvalues which are then given by

$$E_{n\ell}^{\pm} = \frac{w^2\gamma}{2}\left(2n+\ell+\frac{3}{2}\right)^2 \pm \frac{\left(2n+\ell+\frac{3}{2}\right)w}{2}\sqrt{w^2\gamma^2\left(2n+\ell+\frac{3}{2}\right)^2+4} \quad . \tag{43}$$

The requirement of normalizable wavefunction imposes discarding the negative roots. Further, as discussed earlier [17-21], a coherent model is met only for $\gamma \prec 0$. The results in (41) and (43) are in agreement with those in Refs. [18], [20] and [21], for which it is reminded that the principal quantum number $n_p$ is related to the radial quantum number ($n = 0,1,2,...$) used here as $n_p = n+\ell+1$. Finally, we note that the solutions in (41) and (43) reduce explicitly to those concerning with the conventional harmonic oscillator potential in case $\gamma = 0$ which serves as a testing ground.

### 3.3.2. Coulomb potential

This potential is given by

$$V(x, E_{n\ell}) = V_{ES}(x) + \Delta V(x, E_{n\ell}) = \left(\frac{\lambda}{x} + \frac{\ell(\ell+1)}{x^2}\right) + \frac{\lambda}{x}(\gamma E_{n\ell}) \quad , \quad \lambda \prec 0 \quad , \tag{44}$$

as having a negative strength, it requires $\gamma \succ 0$. The analytically solvable energy independent piece of the Coulomb potential has the known solutions

$$E_{ES} = -\frac{\lambda^2}{4(n+\ell+1)^2} \quad , \quad \Psi_{ES}(x) = f(x)F(g) \sim x^{\ell+1} \exp\left[\frac{\lambda}{2(n+\ell+1)}x\right] L_n^{2\ell+1}(g) \quad , \tag{45}$$

for which we choose

$$Q(g) = 0 \quad , \quad R(g) = \frac{n+\ell+1}{g} - \frac{\ell(\ell+1)}{g^2} - \frac{1}{4} \quad , \tag{46}$$

that leads to $g(x) = -\lambda x/(n+\ell+1)$.

For the calculations of the modifying terms, which are finally added to the energy and the reduced wavefunction given in (45) due to the energy dependent part of the potential ($\Delta V$), we set $\Delta Q = -\gamma E_{n\ell}$ and the use of which into Eq. (12), together with the consideration of (14), reproduces

$$\Delta E = -\frac{\lambda^2 \gamma E_{n\ell}}{2(n+\ell+1)^2}\left(1 + \frac{\gamma E_{n\ell}}{2}\right) \quad , \tag{47}$$

and remembering that $\Delta Q = 2h'/hg'$, from Eq. (10), we arrive at

$$h(x, E_{n\ell}) = \exp\left[\frac{\gamma E_{n\ell} \lambda}{2(n+\ell+1)} x\right] \quad . \tag{48}$$

Consequently, the whole of the actual solututions are

$$\Psi(x, E_{n\ell}) = f(x) F(g) h(x, E_{n\ell}) \sim x^{\ell+1} \exp\left[\frac{\lambda x}{2(n+\ell+1)}(1 + \gamma E_{n\ell})\right] L_n^{2\ell+1}(g) \tag{49}$$

where $g = -\lambda(1+\gamma E_{n\ell})x/(n+\ell+1)$ in this case and the sum of the two different energy contributons, $E_{n\ell} = E_{ES} + \Delta E$, gives the energy eigenvalues as the solutions of a second order equation with two roots

$$E_{n\ell}^{\pm} = \frac{1}{\gamma^2 \lambda^2}\left[-2(n+\ell+1)^2 - \gamma\lambda^2 \pm 2(n+\ell+1)\sqrt{(n+\ell+1)^2 + \gamma\lambda^2}\right] \quad , \tag{50}$$

where $E_{n\ell}^{+}$ is the physically acceptable one. The results in Eqs. (49) and (50) agree with those of Ref. [21]. The above expression can also be simplified as

$$E_{n\ell}^{+} = -\frac{1}{\gamma + \frac{2}{\lambda^2}(n+\ell+1)\left[(n+\ell+1) + \sqrt{(n+\ell+1)^2 + \gamma\lambda^2}\right]} \tag{51}$$

which clearly justifies the reliability of (50) due to the reduce of Eq. (51) to the usual energy expression in (45) for the case $\gamma = 0$.

## 4. CONCLUDING REMARKS

An attempt has been made to generalize the work in [5] and shown that the mathematically rigorous new scheme unifies different theories for the solution of Schrödinger equation with analytically/approximately solvable conventional and energy-dependent potentials. The presented algorithm is also found to be equivalent to the alternative model reported previously [16]. This remarkable coincidence has revealed the bridge between the algebraic approach in the scenario introduced in this work and the one carried out within the frame of an extended SUSYQM theory [16], completing the discussion of Lévai [5] regarding the connection between the simple prescription used in his work and the procedure within the usual SUSYQM theory. Although the literature covered similar problems, to our knowledge an investigation such as the one presented here was missing.

In addition, the procedure used here for approximately solvable potentials is well adapted to the use of software systems such as Mathematica and allows the computation to be carried out up to high orders of the perturbation. To go beyond qualitative aspects, the second part of the applications is devoted also to the study of the wave equation with potentials depending on the energy which is essential in understanding the interaction mechanism in heavy quark systems. It has been clarified that such investigations can also be performed safely through our schematical model without causing any physical problem. Although we have limited ourselves to two illustrative examples, the range of application of the method is rather large and appears to be straightforward.

Beyond its intrinsic importance as a new solution for a fundamental equation in physics, we expect that the present simple method would find a widespread application in the study of different quantum mechanical systems with constant and position-dependent masses. In particular, the present discussion would be useful in perturbational treatments of the exact spectra of a few particle systems, and thus provide a further insight on discussion of the fractional nature of such systems. Finally, the remaining question here is to know if the scenario put forward in the present work is applicable to non-central potentials and also, after some necessary modifications, to the related problems in the relativistic region, within the consideration of Eq. 7. Along this line the works are in progress.

| $\mu$ | $\sigma$ | $\eta$ | $N=4$ | $N=8$ | $N=12$ | Exact |
|---|---|---|---|---|---|---|
| 1 | 0 | 0.1 | 1.104 923 | 1.109 628 | 1.109 070 | 1.109 087 |
| | | | 3.576 125 | 3.598 684 | 3.595 729 | 3.596 037 |
| | | | 6.609 983 | 6.662 450 | 6.655 648 | 6.644 392 |
| | | | 10.391 040 | 10.483 375 | 10.472 339 | 10.237 874 |
| | | 1.0 | 1.418 059 | 1.442 229 | 1.435 465 | 1.435 625 |
| | | | 4.971 886 | 5.051 659 | 5.034 736 | 5.033 396 |
| | | | 9.831 164 | 9.974 381 | 9.958 135 | 9.966 622 |
| | | | 16.219 169 | 16.435 265 | 16.391 053 | 15.989 441 |
| | | 10.0 | 2.174 017 | 2.221 521 | 2.205 998 | 2.205 723 |
| | | | 8.002 447 | 8.156 497 | 8.110 650 | 8.114 843 |
| | | | 16.353 667 | 16.624 921 | 16.587 359 | 16.641 218 |
| | | | 27.537 122 | 27.940 075 | 27.843 302 | 27.155 086 |
| | | 100.0 | 3.665 363 | 3.745 295 | 3.718 101 | 3.716 975 |
| | | | 13.751 708 | 14.023 562 | 13.966 820 | 13.946 207 |
| | | | 28.440 597 | 28.925 950 | 28.863 060 | 28.977 294 |
| | | | 48.230 105 | 48.952 973 | 48.770 486 | 47.564 985 |
| | | 1000.0 | 6.404 635 | 6.542 058 | 6.487 758 | 6.492 350 |
| | | | 24.184 202 | 24.664 085 | 24.557 556 | 24.525 316 |
| | | | 50.214 147 | 51.077 401 | 50.968 447 | 51.182 480 |
| | | | 85.350 546 | 86.638 619 | 86.308 303 | 84.175 584 |
| 0 | 0 | 1 | 1.129 584 | 1.153 559 | 1.143 340 | 1.144 802 |
| | | | 4.278 386 | 4.363 353 | 4.340 883 | 4.338 599 |
| | | | 8.899 753 | 9.053 228 | 9.034 111 | 9.073 085 |
| | | | 15.143 475 | 15.372 717 | 15.313 502 | 14.935 169 |

**Table 1.** Comparison of the first four eigenvalues of the potential $\mu x^2 + \eta x^6$ obtained by the present method with the exact values ( [Ref. 24] for $\mu=1$, and Ref. [25] for $\mu=0$ )

| Average SWKB [Ref. 26] | Modified Hill Determinant Method [Ref. 27] | Present Calculations |
|---|---|---|
| 7.3786 | 7.3569 | 7.3569 |
| 24.6861 | 24.6462 | 24.6462 |
| 46.3690 | 46.3355 | 46.3585 |
| 71.3823 | 71.3534 | 73.0669 |

**Table 2.** Comparison of the present calculation results for the first four eigenvalues of the potential $\mu x^2 + \sigma x^4 + \eta x^6$, where $\mu=30$, $\sigma=20\sqrt{30}$ and $\eta=100$, with those obtained with the two different algebraic models